\documentclass[12pt]{article}
 \usepackage{ifpdf}
 \usepackage{epsfig}
  \usepackage[latin1]{inputenc}
 \usepackage{amsfonts}
 \usepackage{amsthm}
 \usepackage{amssymb, latexsym, amsmath}
 \usepackage{amsbsy}
\usepackage{amstext}
\usepackage{amsopn}
\usepackage{amsgen}
\usepackage[dvips]{color}
 \usepackage{graphicx}
  \usepackage{color}
   \usepackage{wasysym}
   \usepackage{deluxetable}
  \usepackage{verbatim}
  \usepackage[T1]{fontenc}

 \newcommand{\be}{\begin{equation}}
 \newcommand{\ee}{\end{equation}}
 \newcommand{\ba}{\begin{eqnarray}}
 \newcommand{\ea}{\end{eqnarray}}
 \newcommand{\bs}{\begin{subequations}}
 \newcommand{\es}{\end{subequations}}

 \newcommand{\erbold}{\mbox{{\boldmath $
 r$}}}

\begin{document}
  \title{
         ${{~~~~~~~~~~~~~~~~~~~~}^{^{^{
       {\rm{   Published~in\,  }}
                    ~Icarus
     \rm   \,,~Vol.~300\,,~pp.~223 - 226\,~(2018)
                  }}}}$\\
 {\Large{\textbf{{{Tidal viscosity of Enceladus}
 ~\\
 ~\\}
            }}}}
 \author{
                                     {\Large{Michael Efroimsky}}\\
                                     {\small{US Naval Observatory, Washington DC 20392}}\\
                                     {\small{e-mail: ~michael.efroimsky$\,$@$\,$navy.mil~$\,$}}\\
                                     ~\\
 }
     \date{}

 \maketitle

 \begin{abstract}
 ~\\
 In Efroimsky (2018), we derived an expression for the tidal dissipation rate in a homogeneous near-spherical Maxwell body librating in longitude.
 %
 %
 %
 Now, by equating this expression to the outgoing energy flux due to the vapour plumes, we estimate the mean tidal viscosity of Enceladus, under the assumption that the Enceladean mantle behaviour is Maxwell. This method yields a value of $\,0.24\times 10^{14}\;\mbox{Pa~s}\,$ for the mean tidal viscosity, which is very close to the viscosity of ice near the melting point.

 \end{abstract}

 \section{Motivation and plan\label{section1}}

 The origin of Enceladus is subject to debate and hypotheses.

 An older scenario suggested that Enceladus and the other regular moons of Saturn completed their accretion concurrently with the planet, about 4.5 Gyr ago (Canup \& Ward 2006). Denser than regular ice, Enceladus must contain a noticeable fraction of rock and metal. It is, however, believed that most of the radioisotope makeup of this fraction was short-lived and by now is gone.$\,$\footnote{~While in the course of their accretion the regular moons might have swept a large number of calcium-aluminum-rich inclusions (CAI's), the radioactive isotopes initially contained in these inclusions~---~mainly $^{26}$Al, plus a small amount of $^{60}$Fe~---~have by now decayed (their half-lives being 0.7 and 2.6 Myr, correspondingly).} Also, owing to the small size of Enceladus, any heat generated in the past by the hypothetical radioactive decay (either short- or long-lived) has long since been conducted away and cannot contribute to the present-day energy budget.
 This way, although radiogenic heating may have played a role in the early development of Enceladus, presently it may be neglected.
 The question then becomes if the tidal heating, enhanced by the forced physical libration in longitude, is sufficient to sustain the persistent plume activity observed around the four large fractures (the `tiger stripes') located near the Enceladean south pole.

 Several recently developed scenarios advocate for a later formation. Charnoz et al. (2011) and Salmon \& Canup (2017) proposed two different accretion processes from the rings, while Asphaug \& Reufer (2013) explored the possibility of a collisional origin of the Saturnian middle-sized moons from large progenitors. Within all these theories, the age of Enceladus gets reduced to about $\,1$ Gyr~---~in which case Enceladus was devoid of radioactive elements since its birth. This again brings up the question whether or not the plume activity can be sustained by tides alone.

 Assuming that Enceladus's mantle behaves as a Maxwell body, we calculate the tidally generated heat (with an input from libration taken into account), equate this heat to the energy emitted by the plumes, and get an estimate for the mean tidal viscosity of Enceladus. The obtained result, $\,0.24\times 10^{14}\;\mbox{Pa~s}\,$, is remarkably close to the viscosity of ice near the melting point. This close coincidence should not be interpreted too literally, given the complex actual structure of Enceladus. Despite this, our method renders a reasonably good approximation, because in calculations of dissipation the presence of a global ocean may be neglected.\,\footnote{~As was shown by Chen, Nimmo and Glatzmaier (2014, Table 3), the dissipation rate associated with the Enceladean global ocean is $\,23.2$ kW , a tiny amount as compared to the overall heat emitted by this satellite.\label{footnote}} ~This way, while the calculation performed for a uniform Maxwell body is obviously approximate, it serves the purpose of showing that the tidal dissipation can alone account for the plumes.

 A separate question, which emerges both within the older and newer scenarios of Enceladus' formation, is whether or not this moon could have been heated up from a cold state and differentiated by tidal heating \,{\it{solely}}. A cold nascent Enceladus had a viscosity much higher than today, wherefore the rate of tidal damping in Enceladus ought, seemingly, to be much lower than now.\,\footnote{~Schubert et al. (2007) explored if the tidal dissipation rate could have been enhanced during a hypothetical initial mild heat-up by long-term radioactivity. While the long-lived radioactivity does not melt the ice in Enceladus, it can warm the ice to the point where tidal heating becomes effective in differentiating the moon. This option, however, becomes less likely within the Enceladus-formation scenarios by Charnoz et al. (2011), Salmon \& Canup (2017), and Asphaug \& Reufer (2013) which set its age as low as 1 Gyr.}
  In reality, however, it could have been quite intense owing to much stronger libration in longitude, had the moon been formed
  with a higher dynamical triaxiality or deformed by a collision.
 This we leave for our next paper (Efroimsky \& Makarov 2018).

 \section{Tidal dissipation rate in the 1:1 spin-orbit resonance}

 Let a planet of a mass $\,M^{\,*}\,$ host a synchronised moon of a mass $\,M\ll M^{\,*}\,$, describing an orbit with a semimajor axis $\,a\,$ and eccentricity $\,e\,$.
 The letter $\,i\,$ will denote the obliquity of the moon's equator on orbit (i.e. the inclination of the planet's apparent orbit as seen from the moon).

 \subsection{The principal modes of forced and free libration in longitude}

 For a synchronised moon, the forced libration in longitude can be approximated with its principal mode (Danby 1962, Frouard \& Efroimsky 2017):
 \begin{equation}
 \begin{split}
 ^{(1:1)}\gamma({\cal{M}})\;\simeq\;
 \bigg[\, -\;6\;e\;\frac{B-A}{C}~\frac{M^*}{M^*+\,M}\;+\;\mathcal{O}(e^3)\,\bigg]\;\sin {\cal{M}}\;+\;\mathcal{O}(e^3)
 \,\;,\;\;\;
 \label{1}
 \end{split}
 \end{equation}
 with $\,A<B<C\,$ being the principal moments of inertia of the moon, $\,e\,$ being the eccentricity, and $\,{\cal{M}}\,$ being the mean anomaly. From the above expression, we see that under synchronism the principal mode has the frequency $\,n\equiv\,\stackrel{\bf\centerdot}{\cal{M}\,}$ and the magnitude
 \ba
 {\cal{A}}_1\;\simeq\;-\;6\;e\;\frac{B-A}{C}\;\,.
 \label{2}
 \ea
  The main frequency of forced libration, $\,n\equiv\,\stackrel{\bf\centerdot}{\cal{M}\,}\,$, coincides with the anomalistic mean motion which can often~\footnote{~For explanation on when such approximation is valid, see  Appendix B to Efroimsky \& Makarov (2014).} be approximated with its Keplerian counterpart $\,\sqrt{G(M+M^{\,*})/a^3\,}\,$.

 Forced libration may be superimposed with free libration. When free libration is weak (less than $\sim 12^{\circ}$), it is sinusoidal and has a frequency $\,\chi\,$ which is extremely low ($\,\chi\ll n\,$) for all planets and almost all moons, Epimetheus being a rare exception.\,\footnote{~It should also be noted that the expressions (\ref{1}) and (\ref{2}) serve as good approximations only for $\,\chi\ll n\,$. See, e.g. equation (21) in Frouard \& Efroimsky (2017).}  Stronger free libration will contain a whole spectrum of modes. We shall assume that free libration is weak, and shall denote its magnitude with $\,{\cal{A}}_{\textstyle{_{\rm{\,free}}}}\,$.

 \subsection{Power exerted by tides, in the presence of libration and obliquity}

 The tidally dissipated power in a satellite consists of several inputs. Besides the leading (``main'') term, which is unrelated to libration or obliquity, the power contains contributions from the forced and free libration, and from the equatorial obliquity on orbit.

 As demonstrated in Efroimsky (2018), these terms sum up to
 \ba
 \nonumber
 ^{(1:1)}\langle P\rangle_{\rm{tides}}\,=\,^{(1:1)}\langle P\rangle_{\rm{tides}}^{\rm{(main)}}\,+\,^{(1:1)}\langle P\rangle_{\rm{tides}}^{\rm{(forced)}}\,+\,^{(1:1)}\langle P\rangle_{\rm{tides}}^{\rm{(free)}}\,+\;^{(1:1)}\langle P\rangle_{\rm{tides}}^{\rm{(obliquity)}}\,=\;\qquad\;\qquad\;\qquad\;\qquad
  ~\\
  \label{b}
  \label{3}\\
  \nonumber
 \frac{G{M^{\,*}}^{\,{2}}\,R^{\,5}}{a^6}\,n\,k_2(n)\;\sin\epsilon_2(n)\,
 \left[\frac{21}{2}\,e^2\,-\,6\,{\cal{A}}_1\,{e}\,+\,\frac{3}{2}\,{\cal{A}}^2_1\,+\,\frac{3}{2}\,{\cal{A}}^2_{\rm{\,free}}\,\frac{\chi_{\textstyle{_{\rm{free}}}}\,k_2(\chi_{\textstyle{_{\rm{free}}}})\;\sin\epsilon_2(\chi_{\textstyle{_{\rm{free}}}})}{n\,k_2(n)\;\sin\epsilon_2(n)}
 \;+\;\frac{3}{2}\;\sin^2 i\right]\,,
 \ea
 where $\,G\,$ is the Newton gravity constant, $\,R\,$ is the radius of the tidally perturbed body (Enceladus, in our case), $\,{M^{\,*}}\,$ is the mass of the perturber (Saturn).
 As ever, $\,k_2\,$ and $\,\epsilon_2\,$ are the quadrupole Love number and phase lag, both being functions of the forcing frequency (see, e.g., Efroimsky 2015). Although expression (\ref{3}) was derived for a homogeneous near-spherical body, application of this expression to Enceladus is justified, as explained in Footnote \ref{footnote}.

 On the left-hand side, the summands $\,P\,$ are put in angular brackets, to denote time averaging, and are equipped with the left superscript \,(1:1)\,, to emphasise that for these summands we use expressions valid for a synchronised moon. In other spin-orbit resonances, the expressions for these summands would be different.

 On the right-hand side, in the square brackets, the first term is ``main'', in that it is related neither to libration nor to the obliquity. (This term is not always the largest, though.) The forced libration with the frequency $\,n\,$ and magnitude $\,{\cal{A}}_1\,$ gives birth to the second and third terms.$\,$\footnote{~Despite the `minus' sign, the second term is positive definite, because in the 1:1 spin-orbit resonance the magnitude $\,{\cal{A}}_1\,$ is negative, see expression (\ref{2}).} The fourth term is owing to the free libration with the frequency $\,\chi_{\textstyle{_{\rm{free}}}}\,$ and magnitude $\,{\cal{A}}_{\rm{\,free}}\,$. The last term is due to the obliquity $\,i\,$ of the satellite's equator on orbit.

 \subsection{Tides in a near-spherical Maxwell body\label{2.3}}

 If we assume that the factor $\,k_2/Q\,=\,k_2\,\sin\epsilon_2\,$ is frequency-independent, the free-libration-produced term in the square brackets becomes  $\;\frac{\textstyle 3}{\textstyle 2}\,{\cal{A}}^2_{\,{{\rm{free}}}}\,\frac{\textstyle \chi_{\textstyle{_{\rm{free}}}}}{\textstyle n}\;$. Then, in the special case of $\,\chi_{\textstyle{_{\rm{free}}}}/n = 1/3\,$, our expression (\ref{b}) will coincide with formula (45) from Wisdom (2004).

 In reality, however, $\,k_2/Q\,=\,{k}_2\,\sin\epsilon_2\,$ is always a function of frequency. For a near-spherical body of an $\,${{\it{arbitrary}}}$\,$ linear rheology, the frequency-dependence of $\,k_2/Q\,=\,{k}_2\,\sin\epsilon_2\,$ has the form of a sharp kink, with its peak located at an extremely low frequency $\,\chi_{peak}\,$. For a near-spherical Maxwell body, that borderline frequency is
  \ba
  \chi_{peak}\,=\;\frac{8\,\pi\, G\,\rho^2\,R^2}{57\,\eta}\,\;,
  \nonumber
  \ea
  $R\,$, $\,\rho\,$ and $\,\eta\,$ being the mean radius, mean density and shear viscosity, correspondingly. To the left of that peak, the function $\,k_2\,\sin\epsilon_2\,$ goes almost linearly through zero, while to the right of the peak it scales (for a Maxwell body) as the inverse frequency:
 \ba
 \frac{{k}_2(\chi)\;\sin\epsilon_2(\chi)}{\frac{\textstyle 2}{\textstyle 3}\;G\;R^{\,2}\;\rho^2}\;\approx\;|\,\bar{J}(\chi)\,|\,\sin\delta(\chi)\;=\;\frac{1}{\eta\,\chi}\,\;,
 \nonumber
 \ea
 where $|\,\bar{J}(\chi)\,|\,\sin\delta(\chi)=\frac{\textstyle 1}{\textstyle \chi\,\eta}\,$ is the negative imaginary part of the Maxwell complex compliance $\,\bar{J}\,$ at the angular frequency $\,\chi\,$. For details, see equations (31) and (67) in Efroimsky (2015).

 Hence, when the body is Maxwell and the frequencies are not too low, the expression $\,\chi\,k_2(\chi)\;\sin\epsilon_2(\chi)\,$ assumes the same value for $\,\chi = n\,$ and $\,\chi=\chi_{\textstyle{_{\rm{free}}}}\;$:
 \ba
 \nonumber
 n\,k_2(n)\;\sin\epsilon_2(n)\,\;=\;
 \chi_{\textstyle{_{\rm{free}}}}\,k_2(\chi_{\textstyle{_{\rm{free}}}})\;\sin\epsilon_2(\chi_{\textstyle{_{\rm{free}}}})\;=\;
 \frac{\textstyle 2}{\textstyle 3}\;G\;R^{\,2}\;\rho^2\;\frac{1}{\eta}
  \ea

 Under these circumstances, the free-libration-caused term in the square brackets in expression (\ref{3}) becomes simply $\,\frac{\textstyle 3}{\textstyle 2}\,{\cal{A}}^2_{\rm{\,free}}\,$, while the overall factor can be written as
 \ba
 \frac{G{M^{\,*}}^{\,{2}}\,R^{\,5}}{a^6}\,n\,k_2(n)\;\sin\epsilon_2(n)\;=\;\frac{\textstyle 2}{\textstyle 3}\,R^{\,7}\,\rho^2\,n^4\,\eta^{-1}\,\;,
 \label{written}
 \ea
  $\rho$ and $\eta$ being the mean density and viscosity, correspondingly. Inserting this into formula (\ref{b}) and moving the multiplier $21e^2/2$ outside the brackets, we write the  power in a form convenient for comparing the libration- and obliquity-caused inputs with the ``main'' input:
 \ba
 ^{(1:1)}\langle P\rangle_{\rm{tides}}\,=\,\frac{7}{\eta}\;R^{\,7}\,\rho^2\,n^4\,e^2
 \left[\,1\,-\;\frac{4}{7}\;\frac{{\cal{A}}_1}{e}\;+\;\frac{1}{7}\;\frac{{\cal{A}}^2_1}{e^2}
 \;+\;\frac{1}{7}\;\frac{{\cal{A}}^2_{\rm{\,free}}}{e^2}
 \;+\;\frac{1}{7}\;\frac{\sin^2 i}{e^2}\,\right]\,\;.\;\qquad\;
 \label{4}
 \ea

 Referring to the ``main'' input, we employ quotation marks, to emphasise that it may be dwarfed by other contributions. For example, estimates in Efroimsky (2018) demonstrate that the obliquity-caused power in the Moon exceeds the ``main'' term by a factor of 1.85$\,$, and thus is responsible for about 65\% of the overall tidal heating. There we also calculated that for some satellites the forced libration in longitude provides a considerable and even leading input into the tidal heating: 52\% in Phobos, 33\% in Mimas, 23\% in Enceladus, and 96\% in Epimetheus. The result obtained for Epimetheus confirms the prediction by Makarov \& Efroimsky (2014) that physical libration can sometimes greatly enhance the tidal dissipation rate.

 \subsection{The tidal power exerted in Enceladus\label{estimate}}

 The orbital parameters and the ensuing tidal quantities for Enceladus are gleaned in Table~\ref{Table1}. We see that the contribution from obliquity may be neglected, while the contribution from the forced libration is 30\% of the main term (i.e., about 23\% of the total tidal power).

 Wisdom (2004) included into the heat budget also an input from a secondary resonance. Associated with a frequency $\,\chi=n/3\,$, that input looked like that from free libration. We shall not consider this effect, and shall set the term with $\,{\cal{A}}_{\rm{\,free}}\,$ in the expression (\ref{4}) equal to zero.

 So only the forced-libration-caused term will be added to the ``main'' term, and the total power dissipated in Enceladus by tides will be:
 \ba
 ^{(1:1)}\langle P\rangle_{\rm{tides}}\,=\,\frac{7}{\eta}\;R^{\,7}\,\rho^2\,n^4\,e^2
 \left[\,1\,+\,0.3\,\right]\;=\;
 \frac{9.1}{\eta}\;R^{\,7}\,\rho^2\,n^4\,e^2\;\approx\;\frac{2.4}{\eta} \times 10^{23}\;\mbox{W}
 \,\;,\;\qquad\;
 \label{5}
 \ea
 where we used the physical and orbital parameters from Table \ref{Table1},
 and where we implied that the viscosity is measured in \,Pa \,s\,.

 {\small
  \begin{table}[htbp]
 \begin{center}
 \begin{tabular}{lccc}
 \hline
 \hline
     parameter                 &   notation   &    units~~    &     values                 \vspace{1mm}\\
 \hline
  \hline
      mass                  &   $M$     &  kg  &        $ 1.08 \times 10^{20}$             \vspace{1mm}\\
 \hline
    mean density                  &   $\rho$     &  kg~m$^{-3}~$  &        $ 1.61 \times 10^3$             \vspace{1mm}\\
 \hline
 mean radius                   &     $R$      &      m         &       $2.52\times 10^5$             \vspace{1mm}\\
 \hline
  mean motion                  &     $n$      & rad~s$^{-1}$  &     $5.31\times 10^{-5}$     \vspace{1mm}\\
  \hline
   semimajor axis              &     $a$      &  m            &     $2.38\times 10^{8}$     \vspace{1mm}\\
  \hline
 eccentricity                  &     $e$      &               &       $0.0045$                 \vspace{1mm}\\
 \hline
 equatorial obliquity
 to orbit                      &     $i$      &               &     $\sim 10^{-6}$          \vspace{1mm}\\
  \hline
 free
 libration
 magnitude                     &${\cal{A}}_{\textstyle{_{\rm{\,free}}}}$&   rad         &        $\qquad 0\;$ /presumably/                \vspace{1mm}\\
 \hline
 forced
 libration
 magnitude                     &${\cal{A}}_1$&   rad         &        $-\;0.0021$                \vspace{1mm}\\
 \hline
 libration-caused input in
   tidal heating
   &  $\,-\,\frac{\textstyle 4}{\textstyle 7}\,\frac{\textstyle {\cal{A}}_1}{\textstyle e}\,+\,\frac{\textstyle 1}{\textstyle 7}\,\frac{\textstyle {\cal{A}}^2_1}{\textstyle e^2}\,$ &  &     $0.30$             \vspace{1mm}\\
\hline
  obliquity-caused
  input in
  tidal heating
  &   $\frac{\textstyle \sin^2i}{\textstyle 7\;e^2}$      &  &    $\sim 10^{-3}$               \vspace{1mm}\\
  \hline
  \hline

 \end{tabular}
 \end{center}
 \caption{\small{. The present-time physical and orbital parameters of Enceladus, and the ensuing libration- and obliquity-generated inputs into the tidal dissipation, as compared to the main input (see equation \ref{4}).
 The value of the main-mode forced-libration magnitude $\,{\cal{A}}_1\,$ is taken from Thomas et al. (2016).
 The value of the obliquity with respect to orbit is borrowed from Baland et al. (2016).}}
 \label{Table1}
 \end{table}
 }

 \section{Enceladus' tidal viscosity}

 An early study of the endogenic activity near Enceladus' south pole (Howett et al. 2011) suggested that the outgoing energy flux due to the vapour plumes might be as high as $\,15.8$ GW. However, a later analysis by the same team (Spencer et al. 2013) resulted in a more conservative estimate, $\,4.7$ GW. Recently, Kamata \& Nimmo (2017) advocated for a higher value of $\,10$ GW:
 \ba
 P_{\rm \,plumes}\,=\;10^{10}\;~\mbox{W}\,\;.
 \label{outplume}
 \ea

 If we assume that this power is of tidal origin, and equate it with the above expression for the tidal power, we shall obtain:
 \ba
 ^{(1:1)}\langle P\rangle_{\rm{tides}}\,=\;P_{\rm \,plumes}\qquad\Longrightarrow\qquad \frac{2.4}{\eta} \times 10^{23}\,=\;10^{10}\,\;.
 \label{}
 \ea
 The viscosity then must assume the value of
 \ba
 \eta\;\approx\;0.24\times 10^{14}\;\,\mbox{Pa~\,s}\,\;,
 \label{eta}
 \label{8}
 \ea
 which coincides with the expected viscosity of the ice shell.$\,$\footnote{~The current convention places the viscosity of the ice shell within the range of $\,10^{13} - 10^{15}$ Pa s which is representative of the viscosity of water ice near the melting point
 -- see, e.g., B$\check{\rm{e}}$hounková et al. (2015, Section `Methods') or
 Barr \& Showman (2009, after eqn 15) who obtained this range of viscosity values  from the
 experimental data provided by Goldsby \& Kohlstedt (2001). As Goldsby \& Kohlstedt (2001) were unable to directly observe diffusion creep even at the smallest grain sizes, they had to describe it based on other diffusivity measurements~---~see their equation (4) and Table 6.
 } On the one hand, this coincidence should not be taken too literally, given the complexity of Enceladus' actual structure (see, e.g., $\check{\rm{C}}$adek et al. (2016) and references therein). On the other hand, the smallness of dissipation in the ocean (see Footnote \ref{footnote}) supports our approach.
 One way or another, the obtained coincidence shows that our back-of-the-envelope estimate was reasonable, and we can deduce from it that the tidal friction inside Enceladus is sufficient, alone, to sustain the vapour plumes.

 Another good thing about the obtained value of viscosity is that it $\,${\it{a posteriori}}$\,$ justifies our employment of the expression (\ref{written}).$\,$\footnote{~Recall that expression (\ref{written})~---~and therefore all our subsequent analysis~---~is valid provided the forcing frequency is located to the right of the peak of the frequency-dependence $\,k_2(\chi)\,\sin\epsilon_2(\chi)\,$. Under synchronism, this implies (see Section \ref{2.3})\,:
 \ba
 n\;>\;\chi_{\textstyle{_{\rm{peak}}}}\,=\;\frac{8\,\pi\,G\,\rho^2\,R^2}{57\,\eta}\quad\mbox{or, equivalently:}\quad\eta\;>\;\frac{8\,\pi\,G\,\rho^2\,R^2}{57\,n}\,\;.
 \nonumber
 \ea
 With the numbers from Table \ref{Table1}, this yields:
 $\;
 \eta\;>\;10^{11}\;\mbox{Pa~s}\;,
 $
 an inequality satisfied well by our estimate (\ref{eta}). } It should also be mentioned that a close value was obtained by Ferraz-Mello et al. (2017) who employed a very different tidal model.

 \section{Conclusions}

  Forced libration in longitude produces about 23\% of the tidal dissipation in Enceladus. Assuming that the Enceladean mantle behaves as a Maxwell body, we have calculated the tidal dissipation rate (with the libration-generated input included) and equated it to the outgoing energy flux due to the vapour plumes erupting from the `tiger stripes'. This enabled us to estimate the mean tidal viscosity of Enceladus. The obtained value, $\,\eta\,=\,0.24\times 10^{14}\;\mbox{Pa~s}\,$, is very close to the viscosity of ice near the melting point.

  Naturally, the mean viscosity is much lower than the viscosity of the surface ice, $\,10^{17} - 10^{19}$ Pa s, measured through observation of the crater islands evolution (Leonard et al. 2017).\,\footnote{~Such a high value of the surface-ice viscosity explains why, despite the low average viscosity, Enceladus is not in a hydrostatic equilibrium. Given the smallness of Enceladus, even a thin rigid outer layer will prevent hydrostatic equilibrium from being achieved.}

  In future, this work should be extended to more detailed rheologies, like the combined Maxwell-Andrade model (Castillo-Rogez et al. 2011).

  \section*{Acknowledgments}

 The author is indebted to Julie C. Castillo-Rogez and John R. Spencer for consultation. The author's special thanks go to the Associate Editor, Francis Nimmo, for our extremely useful conversations on the topic of this work, and for his numerous specific comments and criticisms which proved to be most helpful.

 ~\\
 \noindent
 This research has made use of NASA's Astrophysics Data System.

 \end{document}